\documentclass[pdflatex,sn-mathphys-num]{sn-jnl}

\usepackage{graphicx}%
\usepackage{multirow}%
\usepackage{amsmath,amssymb,amsfonts}%
\usepackage{amsthm}%
\usepackage{mathrsfs}%
\usepackage[title]{appendix}%
\usepackage{xcolor}%
\usepackage{textcomp}%
\usepackage{manyfoot}%
\usepackage{booktabs}%
\usepackage{algorithm}%
\usepackage{algorithmicx}%
\usepackage{algpseudocode}%
\usepackage{listings}%
\usepackage{rotating}%
\usepackage{array}%

\theoremstyle{thmstyleone}%
\theoremstyle{thmstyletwo}%

\theoremstyle{thmstylethree}%

\begin{document}

\title[Article Title]{Text-guided multi-stage cross-perception network for medical image segmentation}


\author[1]{\fnm{gaoyu} \sur{chen}}\email{zy2421102@buaa.edu.cn}

\author*[1]{\fnm{haixia} \sur{pan}}\email{haixiapan@buaa.edu.cn}
\equalcont{These authors contributed equally to this work.}

\affil[1]{\orgdiv{college of software}, \orgname{Beihang University}}




\abstract{Medical image segmentation plays a crucial role in clinical medicine, serving as a key tool for auxiliary diagnosis, treatment planning, and disease monitoring. However, traditional segmentation methods such as U-Net are often limited by weak semantic expression of target regions, which stems from insufficient generalization and a lack of interactivity. Incorporating text prompts offers a promising avenue to more accurately pinpoint lesion locations, yet existing text-guided methods are still hindered by insufficient cross-modal interaction and inadequate cross-modal feature representation. To address these challenges, we propose the Text-guided Multi-stage Cross-perception network (TMC). TMC incorporates a Multi-stage Cross-attention Module (MCM) to enhance the model?s understanding of fine-grained semantic details and a Multi-stage Alignment Loss (MA Loss) to improve the consistency of cross-modal semantics across different feature levels. Experimental results on three public datasets (QaTa-COV19, MosMedData, and Duke-Breast-Cancer-MRI) demonstrate the superior performance of TMC, achieving Dice scores of 84.65\%, 78.39\%, and 88.09\%, respectively, and consistently outperforming both U-Net-based networks and existing text-guided methods.}




\maketitle

\section{Introduction}\label{sec1}

Medical image segmentation, which partitions regions of interest in medical images according to specific anatomical or pathological criteria, is a crucial step in medical image processing and analysis\cite{withey2008review}. Accurate segmentation supports clinical tasks such as diagnosis, treatment planning, and disease monitoring across a wide range of applications\cite{Dai2024ReviewOS}. Encoder-decoder networks with skip connections, such as U-Net\cite{ronneberger2015u}, UNet++\cite{zhou2018unet++}, and TransUNet\cite{chen2021transunet}, have been widely adopted as strong baselines for medical image segmentation. However, vision-only architectures often struggle to capture heterogeneous lesion appearances and long-range contextual dependencies, which can lead to insufficiently discriminative representations for subtle or low-contrast lesions\cite{milletari2016v}.\par

To alleviate these limitations, recent studies have explored \emph{text-guided} medical image segmentation, where textual annotations provide additional semantic cues to assist the delineation of target regions. For example, LViT\cite{li2023lvit} jointly trains on images and their corresponding textual descriptions and has demonstrated notable performance gains over purely visual baselines. More recently, several text-guided approaches have been proposed to leverage semantic prompts for medical image segmentation. SimTxtSeg\cite{xie2024simtxtseg} utilizes simple textual cues to generate pseudo-labels and introduces a hybrid attention module for cross-modal feature fusion, achieving impressive results under weak supervision. Similarly, ABP\cite{zeng2024abp} proposes an asymmetric bilateral prompting strategy that enhances target localization by jointly refining image and text prompts. While these methods clearly highlight the benefit of incorporating language information, they typically perform cross-modal interaction at a single stage (e.g., bottleneck or output layer), which limits the propagation of textual semantics across hierarchical visual features. Moreover, most existing approaches do not explicitly align intermediate cross-modal representations, which may result in suboptimal consistency between language guidance and visual perception, especially in regions with ambiguous boundaries or low contrast.\par

Motivated by these observations, our goal is to more effectively exploit textual descriptions to guide feature learning throughout the entire segmentation pipeline. To this end, we propose a \textbf{T}ext-guided \textbf{M}ulti-stage \textbf{C}ross-perception network (TMC) that enables dynamic, multi-stage interaction between visual and language features and enforces explicit cross-modal alignment from low- to high-level representations. Concretely, a Multi-stage Cross-attention Module (MCM) is introduced to adaptively adjust attention weights for relevant image regions conditioned on the textual description, thereby enhancing the model's ability to capture fine-grained semantic details. In addition, a Multi-stage Alignment Loss (MA Loss) is designed to encourage consistent cross-modal feature alignment across multiple stages of the visual and language encoders, promoting stronger semantic matching between image and text.\par

Our main contributions are summarized as follows:\par
\noindent\textbullet{} We propose a Text-guided Multi-stage Cross-perception network (TMC) that integrates text guidance into medical image segmentation through hierarchical, multi-stage cross-modal interaction, aiming to address the limitations of vision-only and single-stage text-guided methods.\par
\noindent\textbullet{} We introduce a Multi-stage Cross-attention Module (MCM) to facilitate dynamic interaction between image and text features at multiple levels, enabling the model to more accurately localize target regions by leveraging text prompts.\par
\noindent\textbullet{} We design a Multi-stage Alignment Loss (MA Loss) that explicitly enhances the consistency of cross-modal features across stages, thereby promoting robust semantic alignment between visual and language representations.\par
\noindent\textbullet{} Extensive experiments on three publicly available medical image datasets demonstrate that each component of TMC contributes to performance gains and that the proposed method consistently outperforms existing state-of-the-art approaches.\par

\section{Related Works}\label{sec2}

\subsection{Medical Image Segmentation}
With the continuous increase in computational power and the scale of available labeled data, deep learning--based methods have rapidly developed and become the dominant paradigm in medical image analysis, especially for medical image segmentation. Long et al.\cite{long2015fully} first proposed the Fully Convolutional Network (FCN) for image segmentation, which replaces fully connected layers with convolutional layers and enables end-to-end dense prediction. In the same year, Ronneberger et al.\cite{ronneberger2015u} introduced U-Net for medical image segmentation, which has become a fundamental baseline in this field. U-Net employs a symmetric U-shaped encoder--decoder architecture and fuses shallow detail features from the encoder with deep semantic features from the decoder through skip connections. To further enhance feature aggregation and refine skip connections, Zhou et al.\cite{zhou2018unet++} proposed U-Net++ with densely connected encoder--decoder subnetworks. TransUNet\cite{chen2021transunet} represents an early attempt to integrate Transformer blocks with CNNs for medical image segmentation, thereby improving the modeling of long-range dependencies. Beyond U-Net variants and early hybrid CNN--Transformer models, more recent works have further strengthened medical image segmentation backbones and decoders through advanced attention and multi-scale design, such as U-KAN\cite{Li2025UKan} and EMCAD\cite{Rahman2024Emcad}, as well as Transformer-based architectures inspired by general-purpose models like DeepLab and SegFormer.\par

These vision-only architectures have achieved remarkable progress in a variety of medical segmentation tasks. However, they rely solely on image features to perform segmentation, which can make it challenging to accurately delineate lesion regions that exhibit low contrast, heterogeneous appearance, or ambiguous boundaries. In particular, they do not exploit auxiliary semantic information that is often available in clinical practice, such as textual descriptions or report-level prior knowledge. In this work, we go beyond pure image-based segmentation and leverage text prompts as additional semantic guidance to improve the robustness and accuracy of medical image segmentation.

\subsection{Text-Guided Medical Image Segmentation}
In recent years, textual information has been increasingly exploited as auxiliary semantic guidance to alleviate annotation scarcity and enhance feature discrimination in medical image segmentation\cite{ryu2025vision}. Li et al.\cite{li2023lvit} proposed the LViT model, which adopts a dual-U architecture comprising a U-shaped CNN branch and a U-shaped Vision Transformer (ViT)\cite{dosovitskiy2020image} branch. In LViT, fused image-text features are extracted within the U-shaped ViT branch and then integrated via a pixel-level attention module (PLAM); the text-enhanced image features are subsequently fed into the CNN decoder. Lee et al.\cite{lee2023text} introduced the text encoder of CLIP\cite{radford2021learning}, pre-trained on large-scale image-text pairs, to extract language features, and proposed a text-guided cross-position attention module that merges textual and visual features at the bottleneck stage. Compared with segmentation methods that rely solely on image features, these approaches demonstrate the clear benefit of incorporating language cues and highlight the great potential of text-guided medical image segmentation.\par

More recently, several works have further explored text-driven prompting and cross-modal interaction for medical segmentation. ABP\cite{zeng2024abp} introduces an asymmetric bilateral prompting strategy that leverages textual information to refine both image and text prompts, thereby enhancing target localization. SimTxtSeg\cite{xie2024simtxtseg} proposes a framework that uses simple textual prompts to generate high-quality pseudo-labels and designs a Textual-to-Visual Cue Converter together with a Text-Vision Hybrid Attention module to fuse text and image features, achieving strong performance on colonic polyp and MRI brain tumor segmentation under weak supervision. In parallel, MedCLIP-SAMv2\cite{koleilat2025medclip} moves towards universal text-driven medical image segmentation by combining CLIP-style medical vision--language pretraining with SAM-like segmentation capabilities. Diffusion-based text-guided enhancement networks\cite{dong2024diff} further exploit generative priors to refine segmentation boundaries conditioned on textual descriptions, providing another line of evidence that multimodal guidance can benefit medical image analysis. These methods, together with other prompt- or CLIP-based approaches, clearly show that textual guidance can improve the robustness and generalization of medical image segmentation.\par

Despite this progress, many existing text-guided methods integrate language features at only a limited number of stages (e.g., mainly at the bottleneck or output layer), which restricts the propagation of textual semantics across hierarchical visual representations. Moreover, while several works design sophisticated cross-modal fusion modules, explicit constraints on intermediate image--text feature alignment are still rare, which may lead to suboptimal consistency between language guidance and visual perception, particularly in regions with low contrast or ambiguous boundaries. To address these limitations, we propose a Text-guided Multi-stage Cross-perception network (TMC), which introduces a Multi-stage Cross-attention Module (MCM) to enable dynamic image-text interaction at multiple levels and a Multi-stage Alignment Loss (MA Loss) to explicitly align cross-modal features from low-level to high-level stages.

\subsection{Visual Language Model}
Visual-language models are designed to jointly encode visual and linguistic information, enabling models to comprehend language descriptions related to visual content and to process visual information conditioned on textual cues. Radford et al. proposed CLIP\cite{radford2021learning}, which employs contrastive learning on large-scale image-text pairs to automatically discover associations between images and their corresponding descriptions in a shared embedding space. Ji et al. presented MAP\cite{ji2023map}, a self-supervised framework that leverages unlabeled multimodal data to learn general-purpose multimodal representations, which can be used for both multimodal data generation and as a pre-trained backbone for downstream tasks. Wang et al. proposed MGCA\cite{wang2022multi}, which constructs a multimodal graph to enhance image-text alignment via contrastive learning.\par

Several visual-language models have been specifically tailored for medical imaging. LoVT\cite{zhang2022contrastive} is a text-supervised pretraining method for medical image localization tasks that learns localization-aware representations by contrasting image-region features with report-sentence features. By integrating instance-level image-report contrastive learning with local contrastive learning, LoVT improves the model's ability to capture fine-grained local patterns in medical images. GLoRIA\cite{huang2021gloria} is a global-local representation learning framework that contrasts subregions of medical images with words in paired reports, jointly leveraging global and local representations to obtain context-aware multimodal features. PRIOR\cite{muller2022joint} further introduces a prototypical representation learning framework that combines global and local alignment modules and enhances image-report representation learning through a cross-modal conditional reconstruction module.\par

These visual-language models demonstrate that aligning visual and textual representations at both global and local levels is crucial for learning semantically rich multimodal features. Inspired by these advances, our work adopts a BERT-based text encoder to extract language representations and focuses on designing multi-stage cross-modal interaction and alignment mechanisms tailored for medical image segmentation, rather than performing large-scale vision-language pretraining from scratch.

\section{Method}\label{sec3}
Although traditional medical image segmentation methods represented by U-Net and its variants have achieved remarkable success across numerous tasks, their performance is still constrained by two critical limitations. First, these approaches rely solely on visual features inherent in the images and lack explicit modeling of high-level semantic information. Consequently, when confronted with lesions exhibiting ambiguous boundaries, low contrast, or highly variable morphology (e.g., early-stage COVID-19 infection regions or non-mass enhancement lesions in breast cancer), the model may struggle to accurately delineate target areas, especially under limited training data, leading to degraded generalization capability. Second, conventional architectures offer limited interactivity and interpretability: clinicians cannot dynamically guide the segmentation process using natural language instructions (e.g., "segment only the consolidation in the right upper lobe"), which restricts their practical utility in real-world clinical workflows.\par

To address these limitations, recent studies have begun to incorporate textual prompts to enhance the semantic understanding of segmentation models. However, many existing text-guided approaches still exhibit notable shortcomings. In particular, they often perform cross-modal fusion only once at the deepest layer (i.e., the bottleneck) of the encoder, which prevents linguistic semantics from effectively propagating to shallow, high-resolution features and thus hinders precise localization of fine-grained structures. Moreover, explicit consistency constraints on intermediate cross-modal representations are rarely imposed, so visual and textual embeddings can gradually drift apart during feature evolution, a phenomenon that becomes especially problematic when lesion boundaries are indistinct or textual descriptions are abstract.\par

To overcome these challenges, we propose a Text-guided Multi-stage Cross-perception network (TMC) that adopts a multi-stage interaction paradigm, explicitly specifying how and where language should guide visual perception. Unlike single-fusion strategies, TMC performs cross-modal interaction at multiple hierarchical levels (in our implementation, at stages 2, 3, and 4 of the encoder-decoder backbone). This multi-stage design offers three key advantages:\par
(1) \textbf{Fine-grained guidance:} Shallow-stage fusion allows textual cues (e.g., "spiculated margin") to directly modulate high-resolution features, enhancing boundary precision.\par
(2) \textbf{Semantic grounding:} Deep-stage interaction ensures that global anatomical context (e.g., "right upper lobe") is consistently reflected in high-level representations.\par
(3) \textbf{Hierarchical consistency:} By enforcing cross-modal alignment across scales, the model maintains coherent vision-language correspondence throughout the feature hierarchy, mitigating semantic drift during upsampling.\par

The TMC architecture comprises two synergistic components: (1) a Multi-stage Cross-attention Module (MCM) that enables bidirectional, scale-aware interaction between image and text features at multiple depths; and (2) a Multi-stage Alignment Loss (MA Loss) that explicitly enforces embedding consistency between visual and textual representations at each fusion stage. Together, these components encourage language to guide not only \emph{what} to segment but also \emph{where} and \emph{how} to segment, leading to more accurate, interpretable, and clinician-controllable segmentation in complex medical scenarios.\par

\subsection{Multi-stage Cross-attention Module (MCM)}
The Multi-stage Cross-attention Module (MCM) enhances the model's semantic understanding of target regions by enabling dynamic interaction between image and text features. At each selected stage of the encoder, MCM employs cross-attention to adaptively modulate visual representations according to the textual description, thereby improving the localization of lesion regions. This design not only compensates for the limited semantic expressiveness of purely visual features, but also fully exploits cross-modal information through multi-stage feature fusion, which in turn benefits the overall segmentation accuracy.\par

As illustrated in Fig.~\ref{fig:pic}, MCM operates on a multi-stage visual encoder and a multi-stage language encoder. Given an image-text pair, the image and the corresponding textual description are fed into their respective encoders, and each stage is indexed by $i = 1, 2, \dots, n$. The visual encoder is built upon a Swin Transformer backbone, which captures long-range dependencies and outputs a feature map $X_i \in \mathbb{R}^{H_i \times W_i \times C_i}$ at stage $i$. For convenience, we flatten $X_i$ along the spatial dimensions and obtain
\[
V_i \in \mathbb{R}^{N_i \times d_v}, \quad N_i = H_i W_i,
\]
where $d_v$ denotes the visual feature dimension. We append a vision query fusion layer at the end of each selected visual stage (e.g., stages 2, 3, and 4), which consists of a multi-head cross-attention block followed by a feed-forward network to fuse visual and language features. Unless otherwise specified, we apply MCM and the corresponding alignment loss at stages $i \in \{2,3,4\}$, while the first stage is kept purely visual to focus on low-level edge and texture features.\par

For the language branch, we adopt a BERT-base model as a representative Transformer-based text encoder. Given a tokenized textual description, BERT produces a sequence of language features
\[
L \in \mathbb{R}^{T \times d_l},
\]
where $T$ and $d_l$ denote the text length and language feature dimension, respectively. To interact with different visual stages, we project $L$ into stage-specific language features $L_i \in \mathbb{R}^{T \times d}$ via a linear layer, where $d$ is the common embedding dimension used for cross-attention. The first token of $L$ is the \texttt{[CLS]} token, which aggregates global information over all language tokens; we denote the \texttt{[CLS]} token at stage $i$ as $\mathrm{CLS}_i \in \mathbb{R}^{1 \times d}$. We choose BERT-base as the text encoder because it provides strong linguistic representations for domain-specific vocabularies and can be easily fine-tuned with limited task-specific text, whereas CLIP-style encoders are primarily optimized for generic image-caption pairs. A comparison with alternative text encoders is provided in the ablation study.
\par

The multi-head cross-attention (MHCA) module fuses visual and language features by alternately updating both modalities at each stage, so that each modality is refined under the guidance of the other. Given query $Q$, key $K$, and value $V$, MHCA is defined as
\[
\mathrm{MHCA}(Q, K, V) = \mathrm{Softmax}\!\left(\frac{Q K^{\top}}{\sqrt{d_k}}\right) V,
\]
where $d_k$ is the key dimension. In MCM, we first update the visual features using language-guided cross-attention and then refine the language features conditioned on the updated visual features:
\[
F_V^{i} = \mathrm{MLP}\big(\mathrm{MHCA}(V_i, L_i, L_i)\big),
\]
\[
F_L^{i} = \mathrm{MLP}\big(\mathrm{MHCA}(L_i, F_V^{i}, F_V^{i})\big),
\]
where $\mathrm{MLP}(\cdot)$ denotes a position-wise feed-forward network applied after the attention operation. The enhanced visual features $F_V^{i}$ and language features $F_L^{i}$ are then passed to the next stages of the visual and language encoders, respectively. By performing such bidirectional, stage-wise cross-attention, MCM encourages the visual representation at each scale to be explicitly informed by textual semantics, and vice versa, leading to more discriminative and text-consistent multi-scale features.In practice, we adopt a four-stage encoder--decoder backbone (i.e., $n=4$). Following the Swin-Tiny configuration, the channel dimensions are set to $C_1=96$, $C_2=192$, $C_3=384$, and $C_4=768$, and the corresponding decoder channels are symmetrically configured. The CNN and ViT down-sampling blocks in the segmentation part share the same spatial resolutions as the Swin encoder at each stage.

\begin{figure}[!h]
	\centering
	\includegraphics[width=1\linewidth]{pic.png}
	\caption{Overview of the proposed Text-guided Multi-stage Cross-perception network (TMC). The image is encoded by a Swin-based visual encoder, while the textual description is processed by a BERT-based language encoder. At selected stages, the Multi-stage Cross-attention Module (MCM) performs bidirectional cross-attention between visual features $V_i$ and language features $L_i$, producing cross-modally enhanced features $F_V^{i}$. The Multi-stage Alignment Loss $\mathcal{L}_{\mathrm{align}}^{i}$ is applied before each cross-attention block to align stage-wise visual and textual embeddings. The fused multi-scale features are propagated through a U-shaped CNN decoder to generate the final segmentation mask, supervised by the segmentation loss $\mathcal{L}_{\mathrm{seg}}$.}
	\label{fig:pic}
\end{figure}

\subsection{Multi-stage Alignment Loss}
Existing vision-language segmentation methods typically fuse image and text features only at the deepest layer (i.e., the bottleneck) of the encoder. However, lesions in medical images often exhibit complex morphological, scale, and positional characteristics (e.g., "subpleural micronodules" or "ground-glass opacities surrounding bronchovascular bundles"), where relying solely on high-level semantic information is insufficient for precise localization of fine-grained structures. Although shallow visual features preserve high-resolution spatial details, they lack semantic guidance and are thus more susceptible to noise. Therefore, we argue that textual prompts should guide visual perception across multiple scales, from local textures to global semantics.\par

To this end, our alignment scheme involves low-level to high-level features in cross-modal alignment, so that intermediate representations from the visual and language encoders are more effectively embedded into a shared cross-modal space. We next explain where the alignment is applied to the feature hierarchy and how the alignment loss is computed.\par

For each selected stage $i$, alignment is performed \emph{before} the cross-attention operation, so that the visual and language features are encouraged to be semantically correlated prior to fusion. Specifically, we use the visual feature $V_i$ from the Swin-based encoder and the stage-specific language representation $\mathrm{CLS}_i$ from the BERT-based text encoder. Both $V_i$ and $\mathrm{CLS}_i$ are linearly projected into a common feature space of dimension $D$, yielding
\[
z_V^{i} \in \mathbb{R}^{H_i W_i \times D}, \qquad z_L^{i} \in \mathbb{R}^{1 \times D}.
\]
To obtain a global visual embedding at stage $i$, we apply global average pooling over the spatial dimension of $z_V^{i}$:
\[
\bar{z}_V^{i} = \mathrm{GAP}(z_V^{i}) \in \mathbb{R}^{1 \times D}.
\]
We then compute the cosine similarity between the pooled visual embedding $\bar{z}_V^{i}$ and the language embedding $z_L^{i}$ and feed it into a logistic contrastive loss:
\[
\mathcal{L}_{\mathrm{align}}^{i}
= - \log\big(\sigma(\mathrm{sim}(\bar{z}_V^{i}, z_L^{i}) / \tau_i)\big)
- \log\big(1 - \sigma(\mathrm{sim}(\bar{z}_V^{i}, z_L^{i}) / \tau_i)\big),
\]
where $\mathrm{sim}(\cdot,\cdot)$ denotes cosine similarity, $\tau_i$ is a learnable temperature parameter at stage $i$, and $\sigma(\cdot)$ is the sigmoid function. The overall multi-stage alignment loss is obtained by averaging over all $n$ selected stages:
\[
\mathcal{L}_{\mathrm{align}} = \frac{1}{n} \sum_{i=1}^{n} \mathcal{L}_{\mathrm{align}}^{i}.
\]
Intuitively, this loss encourages the visual representation at each scale to be compatible with the corresponding language embedding, thus enhancing cross-modal consistency. In practice, we combine the alignment loss with the standard segmentation loss $\mathcal{L}_{\mathrm{seg}}$ as
\[
\mathcal{L} = \mathcal{L}_{\mathrm{seg}} + \lambda \mathcal{L}_{\mathrm{align}},
\]
where $\lambda$ is a weighting factor. By enforcing multi-stage alignment in this way, the model is guided to learn cross-modal features that are both semantically meaningful and spatially discriminative.

\subsection{Segmentation Part}
Given the multi-stage cross-modal features produced by MCM, the segmentation part is responsible for integrating them into a U-shaped architecture to generate the final segmentation mask. We adopt a dual-path design with a CNN-based path for dense prediction and a ViT-based path to further propagate text-conditioned information.\par

Let $X_i$ and $Y_i$ denote the visual feature maps at stage $i$ from the CNN and ViT branches, respectively, and let $F_V^{i}$ be the cross-modally enhanced visual feature obtained by MCM at the same stage. We fuse $F_V^{i}$ into both branches in a stage-wise manner. For the encoder (down-sampling) path, the update rule can be expressed as
\[
\tilde{X}_i = \mathrm{CONCAT}(X_i, F_V^{i}), \qquad
X_{i+1} = \mathrm{DownCNN}_i(\tilde{X}_i),
\]
\[
\tilde{Y}_i = \mathrm{CONCAT}(Y_i, F_V^{i}), \qquad
Y_{i+1} = \mathrm{DownViT}_i(\tilde{Y}_i),
\]
where $\mathrm{DownCNN}_i(\cdot)$ and $\mathrm{DownViT}_i(\cdot)$ denote the $i$-th down-sampling blocks in the CNN and ViT branches, respectively, and $\mathrm{CONCAT}(\cdot)$ is channel-wise concatenation.\par

In the decoder (up-sampling) path, we adopt a standard U-shaped structure with skip connections. At each decoding stage, the up-sampled feature is concatenated with the corresponding cross-modal feature $F_V^{i}$ (and optionally with the encoder feature $X_i$) and then processed by an up-sampling CNN block:
\[
Z_{i-1} = \mathrm{UpCNN}_i\big(\mathrm{CONCAT}(Z_i, F_V^{i}, X_i)\big),
\]
where $Z_i$ denotes the decoder feature at stage $i$ and $\mathrm{UpCNN}_i(\cdot)$ is the corresponding up-sampling block. The final decoder feature $Z_0$ is passed through a $1\times1$ convolution followed by a sigmoid activation to produce the segmentation probability map. In this way, cross-modal information injected by MCM is propagated through the entire U-shaped network and directly contributes to pixel-level predictions.

\subsection{Overall Optimization}
For training, we use a standard pixel-wise binary cross-entropy loss as the segmentation objective. Given the ground-truth mask $y_j \in \{0,1\}$ and the predicted probability $p_j \in [0,1]$ for pixel $j$, the segmentation loss is defined as
\[
\mathcal{L}_{\mathrm{seg}}
= -\frac{1}{N}\sum_{j=1}^{N}\big[y_{j}\log(p_{j})+(1-y_{j})\log(1-p_{j})\big],
\]
where $N$ denotes the total number of pixels. The final training objective combines the segmentation loss with the proposed multi-stage alignment loss:
\[
\mathcal{L} = \mathcal{L}_{\mathrm{seg}} + \lambda \mathcal{L}_{\mathrm{align}},
\]
where $\lambda$ is a weighting factor that balances segmentation accuracy and cross-modal alignment. This joint optimization encourages the network to learn features that are simultaneously discriminative for pixel-level segmentation and consistent with the textual guidance across multiple stages.

\section{Experiment}\label{sec4}

\subsection{Datasets}
We conduct experiments on three publicly available medical image datasets spanning different imaging modalities (X-ray, CT, MRI), anatomical sites, and disease conditions. For each dataset, the segmentation target is the lesion or infection region delineated at pixel/voxel level. Crucially, for volumetric datasets (CT and MRI), we treat each patient scan as a single case and ensure that all slices derived from the same 3D volume are assigned exclusively to one of the training, validation, or testing sets. This prevents data leakage due to inter-slice correlation and yields a realistic evaluation of model generalization.\par

\textbf{QaTa-COV19}\cite{degerli2022osegnet} is a large-scale chest X-ray dataset for COVID-19 infection segmentation. It contains 9{,}258 images from 3{,}780 unique patients, all of whom are COVID-19 positive (i.e., no healthy controls). Each image is accompanied by a pixel-level binary mask of infected lung regions. Following Li et al.\cite{li2023lvit}, we use the text annotations they provided, which describe the presence and extent of infection (e.g., "Bilateral pulmonary infection, two infected areas, upper left lung and upper right lung."). Since all cases are diseased, there is no class imbalance with respect to disease presence; however, to avoid patient-level leakage, we split the dataset at the patient level and ensure that no patient appears in more than one subset. We randomly divide the 3{,}780 patients into training, validation, and testing sets with a 7:1:2 ratio (approximately 2{,}646/378/756 patients, corresponding to 6{,}480/926/1{,}852 images, respectively).\par

\textbf{MosMedData}\cite{morozov2020mosmeddata} consists of 1{,}000 non-contrast chest CT scans, equally divided into 500 normal (non-COVID) and 500 abnormal (COVID-19 positive) cases. For the abnormal cases, radiologists further assign a severity grade based on the extent of pulmonary involvement: CT-1 (mild), CT-2 (moderate), CT-3 (severe), and CT-4 (critical). The approximate distribution among the abnormal cases is CT-1 ($\sim$150), CT-2 ($\sim$200), CT-3 ($\sim$100), and CT-4 ($\sim$50). For segmentation, voxel-level masks of infected lung regions are provided for the abnormal scans, while normal scans serve as negative examples with empty masks. In addition, we use textual descriptions curated from radiological reports that indicate the presence and extent of infection (e.g., "Unilateral pulmonary infection, two infected areas, upper left lung."). We perform splitting at the case (volume) level: all slices from a given CT scan are included in the same subset. To maintain balanced distributions, we adopt a stratified split with respect to both disease status (normal vs. abnormal) and severity grade (CT-1-CT-4), and divide the 1{,}000 cases into training, validation, and testing sets with a 7:1:2 ratio (approximately 700/100/200 cases).\par

\textbf{Duke-Breast-Cancer-MRI}\cite{saha2018machine} includes 922 dynamic contrast-enhanced (DCE) breast MRI scans from 922 female breast cancer patients, with one scan per patient. All cases are malignant (no benign or healthy controls). Each case is provided with lesion segmentation masks delineating the primary tumor region, and associated clinical information such as lesion type (e.g., mass, non-mass enhancement) and kinetic curve descriptors. While standardized histopathological grading (e.g., Nottingham grade) is not consistently available for all subjects, these lesion descriptors correlate with malignancy aggressiveness and serve as a proxy for lesion heterogeneity. We use text annotations verified by two board-certified radiologists that describe lesion location, morphology, and enhancement patterns. The dataset is split at the patient level into training, validation, and testing sets with a 7:1:2 ratio (approximately 645/92/185 patients), and we ensure that the proportions of lesion types (mass vs. non-mass enhancement) are roughly preserved across the three subsets.\par

\subsection{Implementation Details}
The proposed method is implemented in PyTorch on Ubuntu 20.04.4 LTS with a single NVIDIA RTX 4090 GPU (24 GB memory). Unless otherwise specified, all experiments are conducted with the same implementation and training pipeline across the three datasets.\par

We use the Adam optimizer with an initial learning rate of $3\times 10^{-4}$ for QaTa-COV19 and Duke-Breast-Cancer-MRI, and $1\times 10^{-3}$ for MosMedData. The learning rate is reduced by a factor of 0.1 if the validation performance plateaus for 10 consecutive epochs. The batch size is set to 8 for all three datasets. We train the network for at most 200 epochs and apply early stopping if the validation Dice coefficient does not improve for 20 consecutive epochs. The segmentation loss $\mathcal{L}_{\mathrm{seg}}$ is the pixel-wise binary cross-entropy defined in Section~3.4, and the final objective $\mathcal{L} = \mathcal{L}_{\mathrm{seg}} + \lambda \mathcal{L}_{\mathrm{align}}$ is minimized with $\lambda$ empirically set to $0.1$ for all datasets.\par

For image preprocessing, all 2D slices or projection images are resized to $224\times224$ pixels and normalized to $[0,1]$. Segmentation masks are binarized (non-zero pixels set to 1) and resampled using nearest-neighbor interpolation during spatial transformations to preserve label integrity. For CT and MRI volumes, we operate on axial slices extracted from the 3D scans, while ensuring that all slices from the same volume belong to the same data split as described in Section~4.1. During training only, we apply geometric data augmentation including random rotation (within $\pm 20^{\circ}$), random $k\times 90^{\circ}$ rotations, and horizontal/vertical flipping to improve robustness. No data augmentation is applied during validation or testing.\par

Text prompts are tokenized and encoded using a pre-trained BERT-base model ("bert-base-uncased"), whose hidden size is 768 and number of layers is 12. The resulting token embeddings are truncated or padded to a fixed length of 10 tokens, yielding a text representation $T \in \mathbb{R}^{10 \times 768}$. A linear projection layer maps $T$ to the common embedding dimension $d$ used by the cross-attention modules. During training, all BERT parameters are fine-tuned jointly with the rest of the network, unless otherwise stated in the ablation study.

\subsection{Evaluation Metrics}
We adopt the Dice coefficient (Dice) and mean Intersection over Union (mIoU) as our primary evaluation metrics. Both metrics are computed on the held-out test set. For volumetric datasets (MosMedData and Duke-Breast-Cancer-MRI), segmentation performance is first computed at the slice level and then averaged over all slices belonging to the same case to obtain a per-case score; the final values reported in the tables are the averages over all test cases. For each method, we run the training process with three different random seeds and report the mean and standard deviation of Dice and mIoU (mean $\pm$ std). Statistical significance between the proposed TMC and each baseline is assessed using a paired $t$-test on per-case Dice scores, and improvements with $p<0.05$ are marked with an asterisk in the tables.

\begin{sidewaystable}[htbp]
	\begin{tabular}{c|c|cc|cc|cc|cc}
		\hline
		\multirow{2}{*}{Method} & \multirow{2}{*}{Text} & Parameters & FLOPs & \multicolumn{2}{c|}{QaTa-Cov19}                                                                                 & \multicolumn{2}{c|}{MosMedData}                                                                                 & \multicolumn{2}{c}{Duke-Breast}                                                                                 \\
		&                                                       &(M)       &(G)                     & Dice(\%)\(\uparrow\)                                   & mIoU (\%)\(\uparrow\)                                  & Dice(\%)\(\uparrow\)                                   & mIoU (\%)\(\uparrow\)                                  & Dice(\%)\(\uparrow\)                                   & mIoU (\%)\(\uparrow\)                                  \\ \hline
		UNet                    & \(\times\)            & 14.8                            & 50.3                       & 79.71\(\pm\)0.37                                       & 69.31\(\pm\)0.41                                       & 74.33\(\pm\)0.53                                       & 62.25\(\pm\)0.47                                       & 86.15\(\pm\)0.63                                       & 78.95\(\pm\)0.59                                       \\
		AttUnet                 & \(\times\)            & 31.4                            & 55.7                       & 77.03\(\pm\)0.43                                       & 65.68\(\pm\)0.39                                       & 74.38\(\pm\)0.47                                       & 62.31\(\pm\)0.52                                       & 86.16\(\pm\)0.71                                       & 78.93\(\pm\)0.57                                       \\
		nnUNet                  & \(\times\)            & 19.1                            & 412.7                      & 76.95\(\pm\)0.41                                       & 65.57\(\pm\)0.37                                       & 74.21\(\pm\)0.49                                       & 62.12\(\pm\)0.38                                       & 85.31\(\pm\)0.69                                       & 77.65\(\pm\)0.55                                       \\
		TransUNet               & \(\times\)            & 105.0                           & 56.7                       & 76.71\(\pm\)0.51                                       & 65.29\(\pm\)0.43                                       & 73.87\(\pm\)0.55                                       & 61.69\(\pm\)0.46                                       & 83.81\(\pm\)0.62                                       & 75.35\(\pm\)0.59                                       \\
		UCTransNet              & \(\times\)            & 31.0                            & 54.7                       & 79.02\(\pm\)0.39                                       & 68.36\(\pm\)0.47                                       & 76.74\(\pm\)0.53                                       & 65.42\(\pm\)0.44                                       & \underline{86.54\(\pm\)0.67} & \underline{79.53\(\pm\)0.61} \\
		UNet++                  & \(\times\)            & 74.5                            & 94.6                       & 76.73\(\pm\)0.41                                       & 65.28\(\pm\)0.37                                       & 69.82\(\pm\)0.58                                       & 56.77\(\pm\)0.49                                       & 85.86\(\pm\)0.71                                       & 78.48\(\pm\)0.56                                       \\
		DeepLabV3+              & \(\times\)            & 39.6                            & 31.4                       & 81.16\(\pm\)0.45                                       & 68.39\(\pm\)0.41                                       & 75.89\(\pm\)0.52                                       & 61.29\(\pm\)0.47                                       & 86.42\(\pm\)0.68                                       & 76.33\(\pm\)0.57                                       \\
		SegFormer               & \(\times\)            & 27.5                            & 11.2                       & 78.16\(\pm\)0.43                                       & 64.22\(\pm\)0.38                                       & 64.73\(\pm\)0.61                                       & 49.89\(\pm\)0.55                                       & 83.89\(\pm\)0.66                                       & 72.48\(\pm\)0.62                                       \\ \hline
		LViT                    & \(\surd\)             & 29.7                            & 54.1                       & \underline{83.47\(\pm\)0.33} & \underline{74.71\(\pm\)0.41} & 75.75\(\pm\)0.51                                       & 64.09\(\pm\)0.44                                       & 85.18\(\pm\)0.61                                       & 77.41\(\pm\)0.58                                       \\
		TGANet                  & \(\surd\)             & 19.8                            & 41.9                       & 79.62\(\pm\)0.39                                       & 70.47\(\pm\)0.47                                       & 71.48\(\pm\)0.53                                       & 58.98\(\pm\)0.49                                       & 85.08\(\pm\)0.65                                       & 76.68\(\pm\)0.61                                       \\
		CLIP                    & \(\surd\)             & 87.0                            & 105.3                      & 76.69\(\pm\)0.42                                       & 65.26\(\pm\)0.37                                       & \underline{77.24\(\pm\)0.48} & \underline{66.09\(\pm\)0.43} & 85.86\(\pm\)0.69                                       & 78.52\(\pm\)0.57                                       \\
		MAP                     & \(\surd\)             & 215.6                           & 102.2                      & 74.44\(\pm\)0.41                                       & 62.32\(\pm\)0.41                                       & 75.26\(\pm\)0.51                                       & 63.46\(\pm\)0.46                                       & 85.55\(\pm\)0.64                                       & 77.99\(\pm\)0.59                                       \\
		MGCA                    & \(\surd\)             & 212.9                           & 102.0                      & 75.16\(\pm\)0.44                                       & 63.26\(\pm\)0.39                                       & 76.12\(\pm\)0.52                                       & 64.55\(\pm\)0.47                                       & 85.14\(\pm\)0.63                                       & 77.33\(\pm\)0.58                                       \\
		ours                    & \(\surd\)             & 242.8                           & 145.7                      & \textbf{84.65\(\pm\)0.18$^{*}$}                        & \textbf{76.14\(\pm\)0.42}                              & \textbf{78.39\(\pm\)0.14}                              & \textbf{67.48\(\pm\)0.26}                              & \textbf{88.09\(\pm\)0.68$^{*}$}                        & \textbf{82.74\(\pm\)0.73}                              \\ \hline
	\end{tabular}
	\caption{Quantitative comparison of different methods on QaTa-COV19, MosMedData, and Duke-Breast-Cancer-MRI. Results are reported as mean Dice and mIoU (in \%) on the test set; for each method, we run three trials with different random seeds and report mean $\pm$ std. Bold numbers indicate the best performance, and underlined numbers indicate the second best. An asterisk (*) denotes statistically significant improvement over the best competing baseline ($p<0.05$).}
	\label{table:table1}
\end{sidewaystable}
\begin{table}[ht]
	\centering
		\begin{tabular}{cccccccc}
			\hline
			\multicolumn{2}{c}{Method} & \multicolumn{2}{c}{QaTa-COV19}  & \multicolumn{2}{c}{MosMedData}     & \multicolumn{2}{c}{Duke-Breast} \\ \hline
			MCM          & \(\mathcal{L}_{align}\)     & Dice(\%)\(\uparrow\)       & mIoU (\%)\(\uparrow\)      & Dice(\%)\(\uparrow\)       & mIoU (\%)\(\uparrow\)      & Dice(\%)\(\uparrow\)       & mIoU (\%)\(\uparrow\)      \\ \hline
			\(\times\)   & \(\times\)  & 83.47\(\pm\)0.33 & 74.71\(\pm\)0.41 & 75.75\(\pm\)0.51                                       & 64.09\(\pm\)0.44                                       & 85.18\(\pm\)0.61                                       & 77.41\(\pm\)0.58          \\
			\(\surd\)    & \(\times\)  & 84.07\(\pm\)0.37 & 75.53\(\pm\)0.41 & 77.21\(\pm\)0.49                                       & 65.89\(\pm\)0.52                                       & 87.15\(\pm\)0.63                                       & 80.27\(\pm\)0.59          \\
			\(\times\)   & \(\surd\)   & 83.91\(\pm\)0.36 & 75.31\(\pm\)0.42 & 76.97\(\pm\)0.48                                       & 65.58\(\pm\)0.47                                       & 86.83\(\pm\)0.62                                       & 79.77\(\pm\)0.58          \\
			\(\surd\)    & \(\surd\)   & \textbf{84.65\(\pm\)0.18} & \textbf{76.14\(\pm\)0.42} & \textbf{78.39\(\pm\)0.14} & \textbf{67.48\(\pm\)0.26} & \textbf{88.09\(\pm\)0.68} & \textbf{82.74\(\pm\)0.73} \\ \hline
		\end{tabular}
		\caption{Ablation study of the proposed Multi-stage Cross-attention Module (MCM) and Multi-stage Alignment Loss ($\mathcal{L}_{\mathrm{align}}$) on QaTa-COV19, MosMedData, and Duke-Breast. Results are reported as mean $\pm$ std Dice and mIoU (in \%) on the test set over three runs.}
		\label{table:table2}
\end{table}
\begin{figure}[!h]
	\centering
	\includegraphics[width=0.9\linewidth]{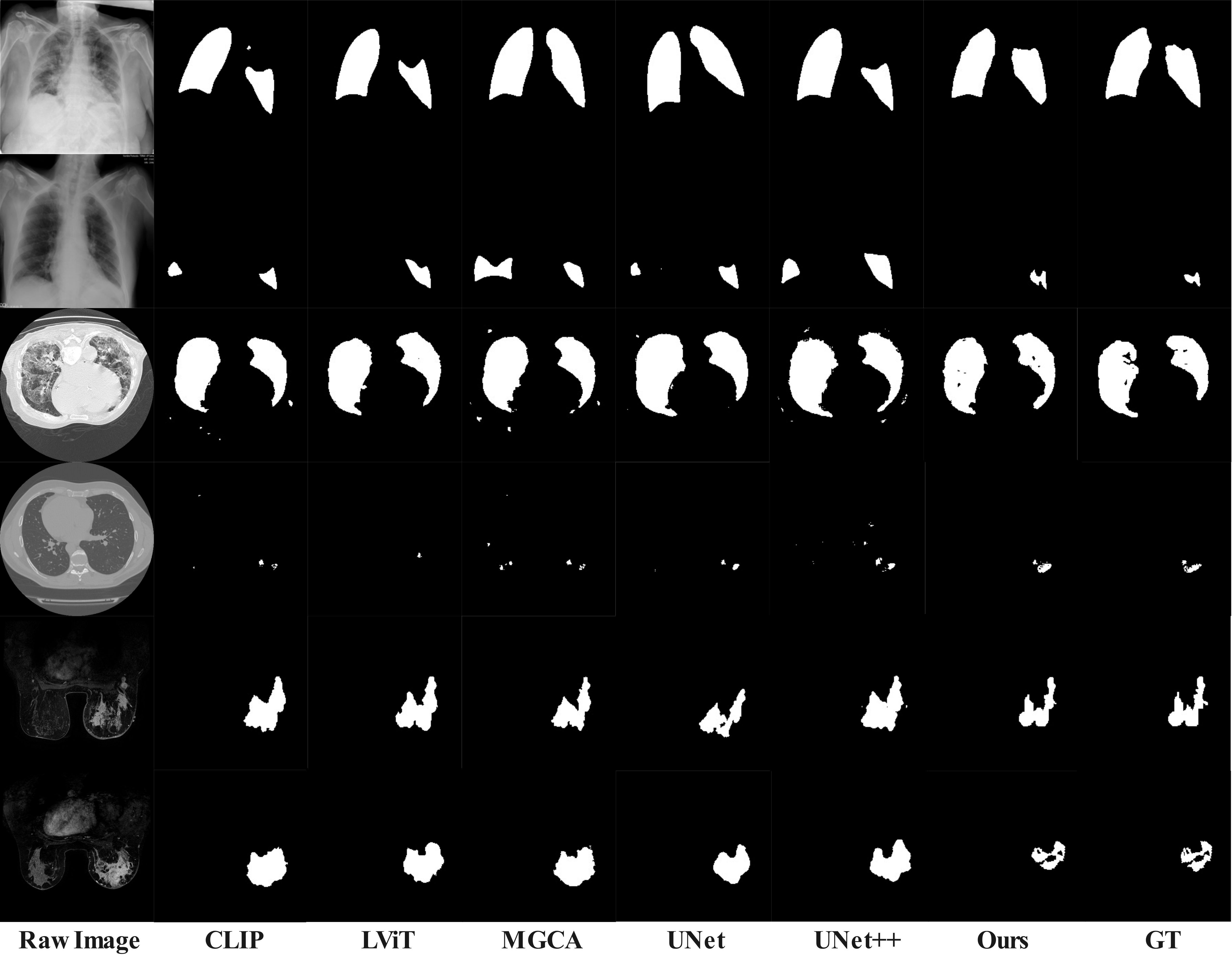}
	\caption{Qualitative comparison of segmentation results on QaTa-COV19, MosMedData, and Duke-Breast-Cancer-MRI. The proposed TMC produces more accurate and sharper lesion delineations compared with state-of-the-art baseline methods.}
	\label{fig:pic3}
\end{figure}
\subsection{Comparison Study}
The TMC is compared with thirteen state-of-the-art segmentation methods, including eight vision-only architectures (U-Net\cite{ronneberger2015u}, AttUNet\cite{oktay2018attention}, nnUNet\cite{isensee2021nnu}, TransUNet\cite{chen2021transunet}, UCTransUNet\cite{wang2022uctransnet}, UNet++\cite{zhou2018unet++}, DeepLabV3+\cite{chen2018deeplab}, SegFormer\cite{xie2021segformer}) and five text-guided or vision-language models (LViT\cite{li2023lvit}, TGANet\cite{tomar2022tganet}, CLIP\cite{radford2021learning}, MAP\cite{ji2023map}, MGCA\cite{wang2022multi}).
\par

\textbf{Qualitative Analysis.}
Fig.~\ref{fig:pic3} illustrates representative qualitative results on QaTa-COV19, MosMedData, and Duke-Breast. For each dataset, we randomly select test cases to visualize typical successes and failure modes. Compared with U-Net and other vision-only baselines, TMC produces more complete coverage of the lesion regions and sharper boundaries, especially for small or low-contrast lesions. In some challenging cases, vision-language baselines such as LViT and CLIP still miss part of the lesion or over-segment surrounding tissues, whereas TMC yields segmentations that are more consistent with the textual descriptions. Nevertheless, as shown in the third column of Fig.~\ref{fig:pic3}, TMC may still under-segment highly ambiguous regions, indicating room for further improvement.\par

\textbf{Quantitative Analysis.}
As summarized in Table~\ref{table:table1}, the proposed TMC achieves the best overall performance on all three datasets. On QaTa-COV19, TMC attains the highest Dice and mIoU scores of 84.65$\pm$0.18\% and 76.14$\pm$0.42\%, respectively, outperforming the strongest vision-only baseline (DeepLabV3+) by approximately +3.5 Dice points and +7.0 mIoU points, and surpassing the best text-guided baseline (LViT) by +1.18 Dice points and +1.43 mIoU points. The Dice improvement over the best competing method is statistically significant ($p<0.05$), as indicated by the asterisk. This suggests that multi-stage cross-perception provides more effective semantic guidance than single-stage fusion or purely visual architectures on chest X-ray infection segmentation.\par

On MosMedData, which involves case-level chest CT segmentation with both normal and abnormal scans across multiple severity grades, TMC also yields the best scores among all methods, with 78.39$\pm$0.14\% Dice and 67.48$\pm$0.26\% mIoU. Compared with the strongest vision-only baseline (UCTransNet), TMC achieves absolute gains of about +1.65 Dice points and +2.06 mIoU points; relative to the best vision?language baseline (CLIP), it improves Dice and mIoU by +1.15 and +1.39 points, respectively. These consistent improvements across both Dice and mIoU demonstrate that multi-stage text guidance and alignment are particularly beneficial in the presence of heterogeneous lesion patterns and varying disease severity.\par

On the Duke-Breast-Cancer-MRI dataset, TMC achieves 88.09$\pm$0.68\% Dice and 82.74$\pm$0.73\% mIoU, clearly outperforming all CNN- and Transformer-based baselines as well as all text-guided competitors. In particular, TMC improves over the best vision-only model (UCTransNet) by +1.55 Dice points and +3.21 mIoU points, and over the best text-guided baseline (CLIP) by +2.23 Dice points and +4.22 mIoU points. The Dice gain over the strongest baseline is statistically significant ($p<0.05$). These results indicate that the proposed multi-stage cross-attention and alignment mechanisms are effective not only for lung infection segmentation but also for challenging breast tumor delineation in DCE-MRI.\par

Although TMC has a larger parameter count and FLOPs than some compact baselines (e.g., SegFormer), the performance gains are substantial and consistent across all three datasets. This highlights a favorable trade-off between accuracy and complexity for clinically relevant segmentation tasks, especially when leveraging rich textual descriptions.

\begin{figure}[!h]
	\centering
	\includegraphics[width=0.75\linewidth]{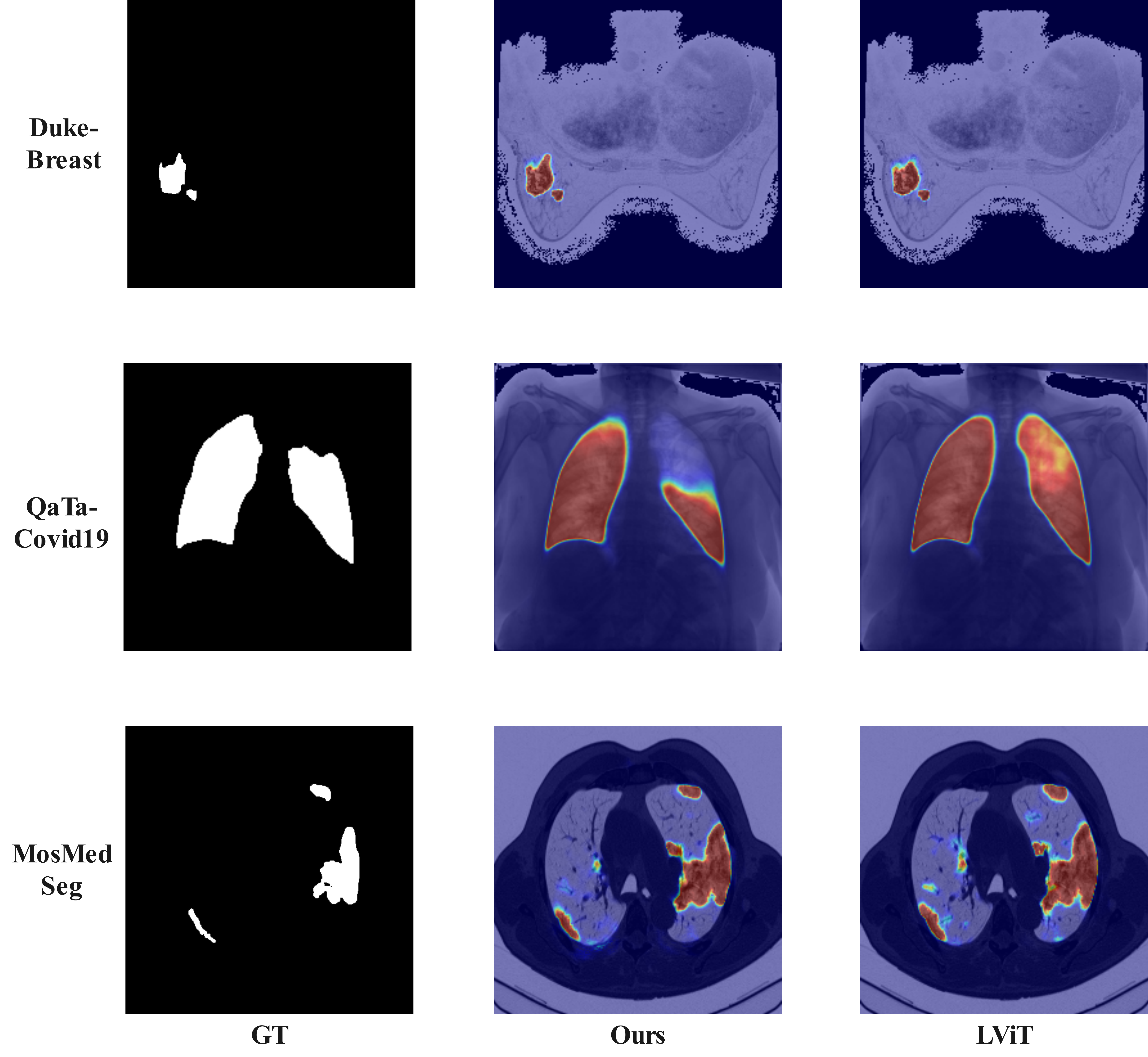}
	\caption{The heatmap demonstrates the strong capability of our TMC in capturing the lesion regions.}
	\label{fig:pic4}
\end{figure}

\subsection{Ablation Study}
We perform ablation experiments to analyze the contribution of the proposed Multi-stage Cross-attention Module (MCM) and Multi-stage Alignment Loss $\mathcal{L}_{\mathrm{align}}$. Unless otherwise specified, all variants share the same backbone and training settings.\par

\textbf{Effectiveness of MCM.}
As shown in Table~\ref{table:table2}, adding MCM alone (\checkmark, $\times$) consistently boosts performance on all three datasets. On QaTa-COV19, Dice and mIoU increase from 83.47$\pm$0.33\% and 74.71$\pm$0.41\% to 84.07$\pm$0.37\% and 75.53$\pm$0.41\%, respectively. On MosMedData, MCM brings larger gains, improving Dice from 75.75$\pm$0.51\% to 77.21$\pm$0.49\% (+1.46 points) and mIoU from 64.09$\pm$0.44\% to 65.89$\pm$0.52\% (+1.80 points). On Duke-Breast, Dice and mIoU increase by +1.97 and +2.86 points (from 85.18$\pm$0.61\% / 77.41$\pm$0.58\% to 87.15$\pm$0.63\% / 80.27$\pm$0.59\%). These improvements confirm that multi-stage cross-attention effectively enhances cross-modal interaction by allowing textual cues to refine visual features at multiple scales.\par

\textbf{Effectiveness of $\mathcal{L}_{\mathrm{align}}$.}
When we remove MCM but retain the multi-stage alignment loss ($\times$, \checkmark), performance also improves over the baseline without either component. On QaTa-COV19, $\mathcal{L}_{\mathrm{align}}$ alone increases Dice from 83.47$\pm$0.33\% to 83.91$\pm$0.36\% and mIoU from 74.71$\pm$0.41\% to 75.31$\pm$0.42\%. On MosMedData, Dice and mIoU improve by +1.22 and +1.49 points (from 75.75$\pm$0.51\%/64.09$\pm$0.44\% to 76.97$\pm$0.48\%/65.58$\pm$0.47\%), while on Duke-Breast they increase by +1.65 and +2.36 points (from 85.18$\pm$0.61\%/77.41$\pm$0.58\% to 86.83$\pm$0.62\%/79.77$\pm$0.58\%). These results indicate that explicitly aligning visual and textual embeddings at multiple stages helps alleviate inadequate cross-modal feature expression, even without explicit cross-attention.\par

\textbf{Full TMC.}
Combining MCM and $\mathcal{L}_{\mathrm{align}}$ (\checkmark, \checkmark) yields the best performance on all datasets. Compared with the baseline without either component, the full TMC improves Dice and mIoU by +1.18 and +1.43 points on QaTa-COV19 (from 83.47\%/74.71\% to 84.65\%/76.14\%), by +2.64 and +3.39 points on MosMedData (from 75.75\%/64.09\% to 78.39\%/67.48\%), and by +2.91 and +5.33 points on Duke-Breast (from 85.18\%/77.41\% to 88.09\%/82.74\%). The consistent gains across three datasets suggest that MCM and $\mathcal{L}_{\mathrm{align}}$ are complementary: MCM provides rich multi-stage cross-modal interaction, while the alignment loss stabilizes training and enforces coherent vision?language correspondence throughout the feature hierarchy.

\subsection{Interpretability Analysis}
Fig.~\ref{fig:pic4} shows heatmap visualizations of the attention maps produced by TMC for representative test cases. The highlighted regions correspond closely to the lesions described in the text prompts (e.g., "subpleural micronodules in the right lower lobe"), suggesting that the model effectively leverages textual guidance to focus on clinically relevant structures. These visualizations provide qualitative evidence for the interpretability of the proposed multi-stage cross-perception mechanism.

\section{Discussion}
In this work, we proposed the Text-guided Multi-stage Cross-perception network (TMC), a framework for medical image segmentation that leverages textual prompts to enhance lesion localization and boundary delineation. Extensive experiments on three diverse public datasets---QaTa-COV19 (chest X-ray), MosMedData (chest CT), and Duke-Breast-Cancer-MRI---demonstrated that TMC consistently outperforms both conventional U-Net-based architectures and recent vision?language segmentation methods. On these datasets, TMC achieves Dice scores of 84.65\%, 78.39\%, and 88.09\%, respectively, together with consistently higher mIoU than all competing baselines. Ablation studies further confirmed the individual and synergistic contributions of the two core components: the Multi-stage Cross-attention Module (MCM) and the Multi-stage Alignment Loss (MA Loss).\par

\subsection{Advancement Over Existing Literature}
Recent efforts in text-guided medical image segmentation, such as LViT\cite{li2023lvit}, CLIP-based approaches\cite{lee2023text}, ABP\cite{zeng2024abp}, and SimTxtSeg\cite{xie2024simtxtseg}, have demonstrated the potential of language prompts to provide semantic guidance beyond pixel-level supervision. However, most existing methods perform cross-modal fusion at a single stage (e.g., at the bottleneck or decoder input), limiting the propagation of textual semantics through the hierarchical visual feature representations. In addition, explicit mechanisms to enforce consistency between intermediate visual and linguistic representations are often absent, which can lead to misalignment, particularly in regions with ambiguous boundaries or subtle pathological patterns.\par

TMC addresses these gaps through a multi-stage cross-perception paradigm. First, MCM enables dynamic, bidirectional interaction between image and text features at multiple encoder-decoder stages, allowing the model to adaptively reweight visual attention based on fine-grained textual cues (e.g., "bilateral ground-glass opacities" or "non-mass enhancement"). Second, MA Loss explicitly aligns low- to high-level cross-modal embeddings via contrastive learning, encouraging semantic meaning to be preserved throughout the network. As visualized in Fig.~\ref{fig:pic4}, this design leads to more precise and context-aware activation maps over lesion regions, which in turn translate into improved segmentation accuracy (Fig.~\ref{fig:pic3}). In this sense, TMC moves from single-point to multi-stage cross-modal reasoning, providing a complementary perspective to existing text-guided segmentation frameworks.\par

\subsection{Clinical Implications}
From a clinical perspective, TMC offers several practical advantages. By accepting natural language descriptions as input prompts, it enables intuitive human-AI interaction, allowing clinicians to guide segmentation using familiar radiological terminology without requiring additional manual pixel-level annotation. This is particularly valuable in resource-constrained settings or emerging disease outbreaks (e.g., early-phase pandemics), where high-quality masks are scarce but textual reports may already exist. Furthermore, the high segmentation fidelity achieved on tasks such as COVID-19 lung infection quantification and breast tumor delineation suggests that TMC could support downstream workflows including treatment response assessment, surgical planning, and longitudinal disease monitoring by providing reliable, reproducible lesion metrics. The case-level evaluation protocol (ensuring no patient overlap across splits) further strengthens the clinical relevance and robustness of the reported performance.\par

Another important aspect is interpretability. The cross-attention maps and heatmaps derived from MCM provide a visual explanation of how textual cues modulate model focus, highlighting lesion regions that correspond to specific phrases in the prompt. Such visualizations can increase clinician trust and facilitate error analysis by revealing when the model over-relies on certain textual or visual patterns.\par

\subsection{Limitations}
Despite its strengths, this study has several limitations.\par
First, performance is contingent on the quality and specificity of input text prompts. Ambiguous, incomplete, or non-standardized clinical descriptions (e.g., "abnormality noted" vs. "spiculated mass in the upper outer quadrant") may degrade segmentation accuracy. Future work could integrate medical ontologies (e.g., RadLex) or large language models to normalize, standardize, and enrich raw radiology reports.\par

Second, all experiments were conducted on retrospective, single-center datasets. Generalizability across multi-institutional settings with variations in imaging protocols, patient demographics, and scanner vendors remains to be systematically evaluated. Incorporating multi-center cohorts and external test sets will be an important next step.\par

Third, TMC currently requires paired image-text data for every training sample, which may limit its applicability in scenarios where textual annotations are incomplete or unavailable. Exploring semi-supervised, weakly supervised, or zero-shot adaptation strategies (e.g., using pseudo-text prompts or report-level supervision) would broaden its utility.\par

Fourth, while we ensured case-level data splitting to avoid slice-level leakage, we did not comprehensively evaluate inference speed, memory footprint, or computational efficiency. The multi-stage cross-attention design introduces additional computation compared with purely visual baselines, and a more detailed analysis of the trade-off between performance gain and efficiency (e.g., via model pruning or lightweight attention blocks) is left for future work.\par

\subsection{Outlook}
In summary, TMC advances the integration of vision and language in medical image analysis by introducing multi-stage cross-attention and alignment mechanisms that improve semantic grounding and segmentation robustness. Looking ahead, we envision extending this framework toward: (1) automated or semi-automated prompt generation from unstructured reports, (2) federated or privacy-preserving training across hospitals to enhance generalizability while protecting patient data, and (3) integration into clinical decision support systems for interactive, language-driven diagnostic assistance. We believe such developments could help accelerate the translation of multimodal AI systems from research prototypes to real-world clinical impact.

\section{Conclusion}
In this paper, we presented TMC, a text-guided medical image segmentation framework that addresses insufficient cross-modal interaction and inadequate cross-modal feature expression through a Multi-stage Cross-attention Module (MCM) and a Multi-stage Alignment Loss (MA Loss). By performing bidirectional image-text interaction at multiple encoder-decoder stages and enforcing stage-wise alignment between visual and linguistic embeddings, TMC strengthens semantic grounding and enables more accurate localization of lesion regions guided by textual descriptions.\par

Extensive experiments on three public datasets (QaTa-COV19, MosMedData, and Duke-Breast-Cancer-MRI) demonstrated that TMC consistently outperforms both U-Net-based architectures and recent vision-language baselines in terms of Dice and mIoU. Ablation studies further confirmed that MCM and MA Loss contribute complementary benefits to cross-modal interaction and feature consistency. In future work, we plan to extend TMC to settings with limited or noisy text annotations, improve computational efficiency, and explore integration into clinical decision support systems for interactive, language-driven image analysis.




\end{document}